\documentstyle[aps,multicol,epsf]{revtex}

\def\mbf(#1){\mbox{\boldmath $#1$}}

\begin{document}

\draft

\preprint{HEP/123-qed}

\title{Zero-Conductance Resonances due to Flux States in Nanographite
Ribbon Junctions} 
\author{Katsunori Wakabayashi$^{1,2}$ 
and Manfred Sigrist$^1$ }
\address{$^1$Yukawa Institute for Theoretical Physics, Kyoto University,
Kyoto 606-8502, Japan}
\address{$^2$Institute of Materials Science, University of Tsukuba, Tsukuba
305-8573, Japan}

\date{\today}

\maketitle

\begin{abstract}
The electronic transport properties through junctions in
nanographite ribbons are 
investigated using the Landauer approach.
In the low-energy regime ribbons with zigzag
boundary have a single conducting channel of
edge states. The conductance as a function of the chemical
potential shows a rich structure with sharp dips of
zero conductance. Each zero-conductance resonance is connected with
a resonant state which can be interpreted as the superposition of 
two degenerate flux states with Kekul\'e-like current patterns.  
These zero-conductance dips are connected with 
a pronounced negative magneto resistance. 
\end{abstract}


\begin{multicols}{2}[]

The discovery of fullerene molecules and carbon nanotubes
has triggered intensive research on 
various nanometer size carbon materials\cite{review1,review2}.
In these systems, the geometry of sp$^2$ carbon networks
crucially affects the electronic states near the Fermi level.
Studies with scanning tunneling microscopy and
spectroscopy have confirmed
the connection between 
the electronic states of the 
single wall carbon nanotubes (SWCN)
and their geometry\cite{stm}.
Recently the electrical transport measurement of  
individual SWCN became possible \cite{mceuen,mceuen2,bezryadin,tans} and
the quantized conductance of multi-wall carbon nanotubes
 was observed \cite{qcon}. This initiated theoretical studies
devoted to effects of
non-magnetic impurities\cite{nakanishi}, 
electron correlation\cite{kane}  and
topological defects\cite{chico,tamura1,matsumura,tamura2,igami}. 
Besides these closed carbon molecules, there are also systems 
with open boundaries which display unusual
features connected with their shape. They include small scale 
systems based on graphite, so-called nanographites. 
There are two basic shapes of regular graphite edges, called 
armchair and zigzag edges, depending on the cutting direction of
the graphite sheet.
Properties originating from such edges have 
been studied recently using the model of graphite ribbons, 
one-dimensional graphite stripes of infinite length and finite
width. Ribbons with zigzag edges (zigzag ribbons) possess 
electron states localized near the edge with energies very close to
the Fermi level  \cite{peculiar,nakada,waka}.  
Such states are absent for ribbons with armchair edges. 
The edge states of zigzag ribbons 
were analyzed in terms of nearest-neighbor tight binding
models \cite{peculiar,nakada,waka} 
and density functional approach\cite{miyamoto}.
It was also pointed out that
the edge states play important roles in the magnetic properties
in nanometer-size systems due to their relatively large contribution
to the density of states at the Fermi energy
\cite{peculiar,waka}. Recently, experimental evidence for edge states
has been reported for nanographite systems derived from graphitized
diamond nanoparticles \cite{oeandersson}.
 
In this letter, we investigate the transport properties of ribbons 
related to these edge states. 
For this purpose we design three different types of
junctions which connect zigzag ribbons of the same or different widths. 
The electronic states are described by a single-orbital
nearest-neighbor tight binding model. The conductance of the 
junctions is evaluated using 
the multi-channel Landauer formula\cite{landauer},
\begin{eqnarray}
G(E) = \frac{{\rm e}^2}{\pi\hbar} \sum_{\mu,\nu}|t_{\mu\nu}(E)|^2,
\end{eqnarray}
\noindent
where $t_{\mu\nu}(E)$ is a transmission  coefficient 
from $\mu$-th channel to $\nu$-th channel at energy $E$, calculated by
a recursive Green's function method \cite{green}. 
Before discussing the design of the junctions and their
conductance properties, we review a few facts
concerning the low-energy states in two kinds of graphite ribbons, 
the zigzag and the ``bearded'' ribbon (a variant of a zigzag ribbon). 

{\it Zigzag ribbon:} The zigzag ribbons are metallic for arbitrary
ribbon width. The most remarkable feature is the presence of
a partly flat band at the Fermi level, where the electrons
are strongly localized near the zigzag edge.
Each edge state has a non-vanishing amplitude
only on one of the two sublattices, i.e. non-bonding character.
However, in a zigzag ribbon of finite width, two edge states coming
from both sides, have finite overlap. Because they are located on
different sublattices, they mix into a bonding and anti-bonding
configuration. In this way the partly flat bands acquire a
dispersion [Fig.1(b)]. Note that the overlap is increasing as $k$
deviates from $ \pi/a $, 
because the
penetration depth of the edge states 
increases and diverges at $ k = 2\pi/3a $,
where $ a $ is the lattice constant.
The dispersion depends on the ribbon width $ N $ (number of
zigzag lines from one side to the other).
Close to $k=\pi/a$, the spectrum has the approximate form
$ E_k = \pm 2tN D_k^{N-1} ( 1 - D_k/2) $,
where $D_k = 2\cos\left(\frac{ka}{2}\right)$, 
and $t$ is the hopping matrix element.
Thus, although the edge states on each side separately have
non-bonding character, 
together through their overlap they provide one conducting channel
except at exactly $ E=0 $.
The energy region of single-channel transport 
is restricted by the energy gap ($\Delta_z$) to the next channel 
$\Delta_z \sim 4t\cos [(N-1)\pi/(2N+1)]$ \cite{waka}.

{\it Bearded ribbon:} The bearded ribbon has one zigzag edge and one
edge which has additional bonds (beard) attached to the
zigzag edge [Fig.1(a) region M]. The edge states
of both sides reside on the same sublattice so that in spite of their
overlap their non-bonding character is retained and a completely flat
band at $ E=0$ is resulting for any width $ N $ [Fig.1(b)]. The absence of 
dispersion leads to the insulating behavior for the edge state
channel. The gap ($\Delta_b$) to the first conducting channel 
is given by $ \Delta_b \sim 4t\cos [N\pi /(2N+2)] $ \cite{waka}.

We now turn to the design of a junction connecting
two zigzag ribbons denoted by  L (left) and  R (right). As a first example
we consider the case depicted in Fig.1(a) where the junction region
denoted by M is a bearded ribbon of length $l$ (number of attached
bonds). This junction model represents a metal-insulator-metal
junction providing an illustrative example for the peculiar transport
properties of graphite ribbons regardless of the question as to
whether bearded ribbons could be realized in nature.  
We calculate the conductance $ G(E) $
numerically within the energy range $ |E| < \Delta_b/2 $. 
The result for
$ N=30 $ and different values of $ l $ is shown
in Fig.1(c) using a logarithmic scale on the energy
axis. The most striking feature is the large number of
zero-conductance dips 
at energy values which depend on $ l $. These dips represent
resonances of total reflection. In the whole energy range 
no transmission resonances are observed 
for any value of $ l $. Note also that the change of $N$ 
does not lead to a qualitative change and only modifies the energy
range of single-channel conductance.

\end{multicols}

\begin{figure}[t]
\epsfxsize=\hsize
\epsffile{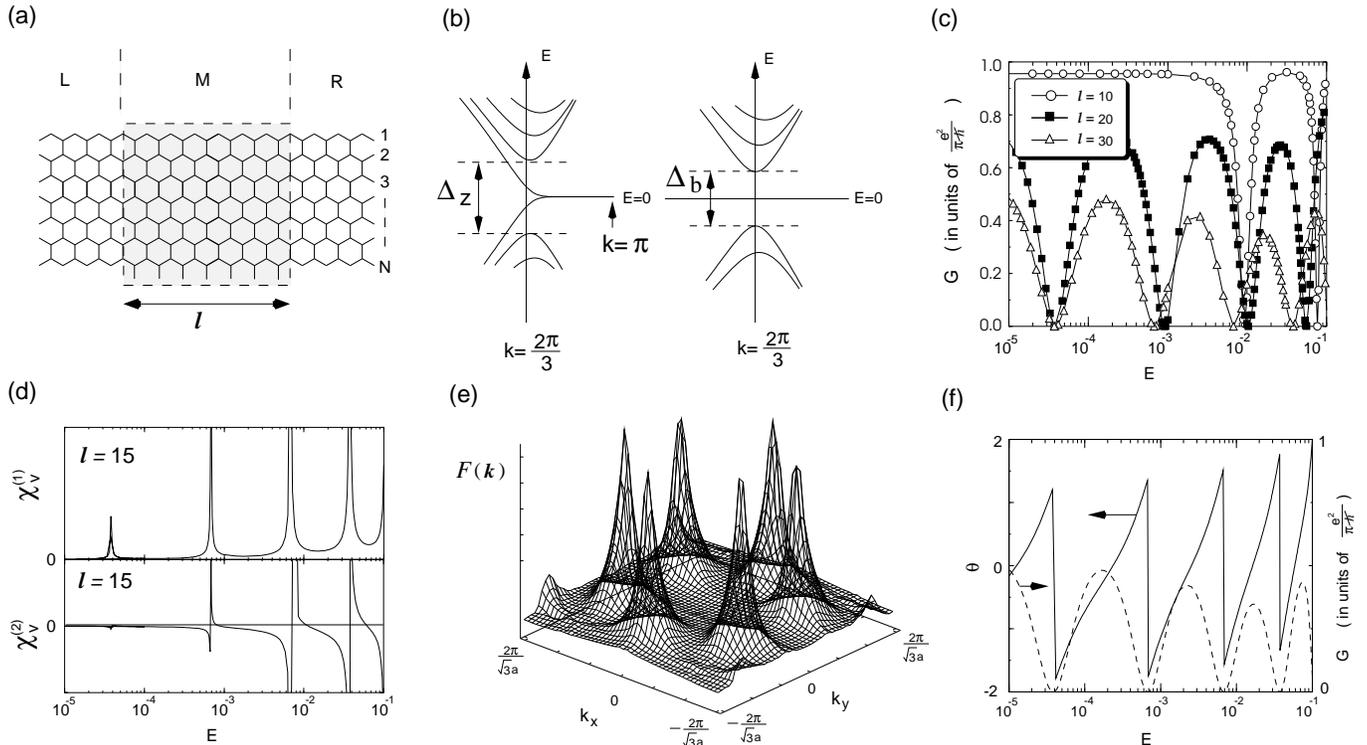}

\caption{(a)The structure of the graphite ribbon junction.
The length of the junction, $l$, is given by the number of attached bonds
in the shaded region.
(b) The schematic figure of energy band dispersion near $E=0$ of
zigzag ribbons (left) and bearded ribbons (right). 
(c) The energy dependence of the conductance, 
for $l=5,10,15$ and $N=30$.
(d) The energy dependence of the response functions,
$\chi_V^{(1)}$ and $\chi_V^{(2)}$
for $l=15$ and $N=30$.
(e) The 3D plot of the correlation of circular current pattern in
the M-region. 
(f) The energy dependence of the phase of the transmission coefficient
of $l=15$ when $N=30$. 
}
\end{figure}

\begin{multicols}{2}[]

Resonances are associated with discrete quantum levels in the junction
regions. In the following we would like to characterize the quantum
states encountered here and discuss the origin of total
reflection. 
It was noticed earlier that some graphite ribbons 
form a triangular Kekul\'e pattern of circular current
driven by special external boundary conditions \cite{waka}. 
To investigate whether the junction states responsible for the
resonance display a similar structure, we introduce the current 
vortex amplitude
defined on the dual (triangular) lattice as the clockwise circular 
sum of the currents $I_{i,p}$ on the bonds of the $p$-th hexagonal
plaquette, $ V_p = \sum_{i=1}^6 I_{i,p}$.
An incident current $ J_{\rm in} $ coming from lead $ L $ 
yields a current flow in the junction region,
whose circulating component (vorticity),
we probe by the following two response functions $ 
\chi_{V}^{(1)} = \langle | V_p | \rangle /J_{\rm in} $ and $
\chi_{V}^{(2)} =  \langle   V_p   \rangle /J_{\rm in} $,
where the bracket $\langle A \rangle$ denotes
the average value of $A$ in the junction region. 
The first function is a measure for the presence of
circular currents and the second for the overall vorticity of the
system. In Fig.1(d) the energy dependence of both functions is
shown for $ l=15 $ and $ N = 30 $. 
The susceptibilities  $ \chi_{V}^{(1,2)} $ show sharp divergences in
the sense of linear response to the driving current $ J_{\rm in} $ when
$ E $ approaches the energy values of a zero-conductance dip. (Note
that $ \chi_{V}^{(1,2)} $ are not defined at zero-conductance energies, 
because no driving current $ J_{\rm in} $ can flow.) This
suggests that the states at 
these energy values form current vortex patterns. The
overall vorticity measured by $ \chi_{V}^{(2)} J_{\rm in} $ 
changes sign at each resonance. 
We find that the circular current form a Kekul\'e-like pattern. This
is most easily seen in the 
Fourier transform 
$ F(\mbf(k)) = \sum_{p } \left(\frac{V_p}{J_{\rm in}} \right){\rm
e}^{i\mbf(k)\cdot\mbf(r)_p}$ in the junction region
which has a clear peak structure of the
triangle correlation of the circular currents pattern (Fig.1(e)) 
($ \mbf(r)_p $: coordinate of the ring center;
$k_x (k_y)$: wave number along (perpendicular to) zigzag lines in the
junction).
This type of pattern appears in $ F(\mbf(k)) $ very close to every
zero-conductance dip and gradually disappears when deviating from
the resonance energies. Because $ F(\mbf(k)) $ is the
correlation among the circular currents on the hexagonal
plaquettes, the state driven by the incident current is in this sense
a flux state. The quantum state associated with the resonance is the
superposition of the two flux states with opposite circular currents,
which combine to a standing wave within the junction region under the
condition of time reversal symmetry. 

The presence of this quantum level plays an important role for the
realization of the zero-conductance resonances. At each resonance the
transmission is not only carried by the usual tunneling through the
insulating junction region, but also through this resonant state. This
decomposition into two channels yields the following form for the
transmission amplitude between the two leads close to a resonance at
energy $ E_0 $,

\begin{equation}
t(E) = \tilde{t}(E) \left( 1 - \frac{i \Gamma/2}{E-E_0 + i \Gamma/2} 
\right) = \frac{\tilde{t}(E)(E-E_0)}{E-E_0 + i \Gamma/2} ,
\end{equation}
where $ \tilde{t}(E) $ is a regular complex function of $ E $ and $
\Gamma $ the width of the resonance. The destructive interference of
the two channels is a consequence of the symmetric form of the $S$-matrix 
which relates the in- and outgoing waves of the two leads with and
of the junction region \cite{shao}. This situation has been discussed in
the context of the three-way splitter and a rigorous proof was given
for the exact cancelling at the resonant energies \cite{shao}. A
consequence of this resonant 
form of the transmission amplitude (Eq.(2)) is that the phase $ \theta
$ of $ t(E) $ exhibits a jump by $\pi$ at each resonance (see Fig.1f)
(similar to Ref. \cite{phase,tani}). The behavior of the phase is
well-described by Eq.(2). 
The form of the $S$-matrix mentioned above as well as the degeneracy
of the flux states of opposite chirality, combining to a 
standing wave state, are based on 
time reversal symmetry. This symmetry can be destroyed by 
applying a magnetic field. We observe a pronounced negative
magnetoresistance at the zero-conductance dip where the conductance 
grows proportionally to $ B^2 $ for small magnetic fields $B$, as we will
discuss in detail elsewhere. 
We would like to mention here that circular currents associated
with zero-conductance dips were recently also reported for
two-dimensional wave guides including junctions with stub geometry
\cite{xu}. Also in this case a negative magnetoresistance should
occur.  

\begin{figure}[t]
\epsfxsize=0.9\hsize
\epsffile{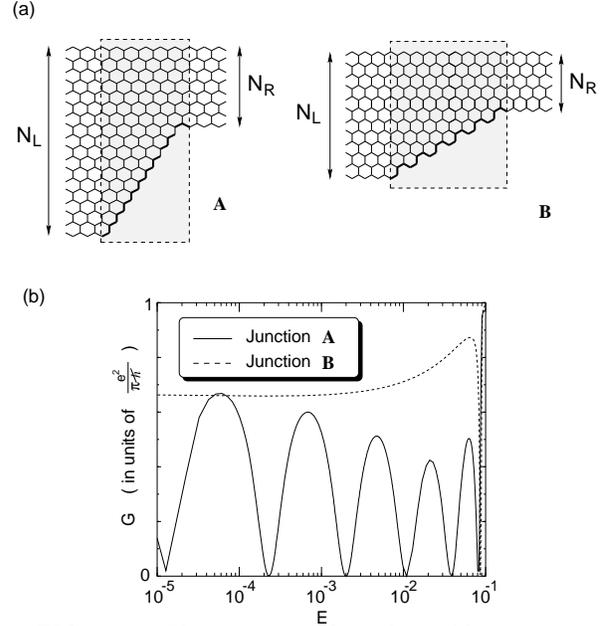}

\narrowtext
\caption{(a) The structure of graphite ribbons junction 
type A and type B.
(b) The energy dependence of the conductance 
for $N_L=50$ and $N_R=30$. }
\end{figure}

Next, we discuss two 
junctions with a more realistic design which do not contain bearded
ribbons. They are shown in Fig.2(a) and connect zigzag ribbon leads of
different width (we show $ N_L=50 $ and $ N_R =30 $ as a
representative case). 
The M-region contains a tilted zigzag edge for
junction A and an armchair edge for junction B. 
The conductance of the two junctions
is qualitatively different. The tilted edge in A supports an edge
state, similar to the bearded ribbon, on the same
sublattice as the edge state on the other side. 
Indeed we find a very similar behavior of the conductance with 
large number of zero-conductance resonances [Fig.2(b)] which are
associated with flux states too.
For junction B there is no localized state near the
armchair edge. The conductance is rather featureless
without any resonance in the energy regime of single-channel
transport. There is a zero-conductance dip above the single-channel 
regime of the lead on the left-hand side ($ E > \Delta_z (N=50)/2 =
0.062 t $) for both 
junctions. The origin is also connected with 
analogous junction states which are based on flux states as
the analysis of the susceptibilities $ \chi_{V}^{(1,2)} $ show,
although the conditions are different here, since the transmission occurs
from three channels on the left hand side to a single channel on the
right hand side. We omit 
here a detailed discussion of this more complicated case. 
For energies above the single channel threshold of the right-hand side 
lead ($ E > \Delta_z (N=30) /2 = 0.101t $) no zero-dip features appear in
any case. 

From this analysis we conclude that the edge structure of the junction 
plays an important role in forming the resonant states in the M-region. 
It is 
well known that the electron states display chiral properties in
graphite sheets, if there
is an imbalance between the two sublattices e.g., by different onsite
potential\cite{somenoff}. The edge states on zigzag edges are a
consequence of this imbalance, 
since the outer most sites belong to a single sublattice. The abrupt
change of the sublattice on the edge as it occurs for previous
junction and junction A yields the boundary condition to form the
degenerate flux states in the M-region. 
These flux states combine to one resonant state of standing-wave nature. 
Further numerical analysis shows that even single non-magnetic
impurities and other simple defect structures disturbing the
sublattice balance can cause a zero-conductance dip associated with
flux states in zigzag ribbons. 

In conclusion, we numerically analyzed three types of nanographite
junctions.  
We found that  the conductance of various junctions having 
zigzag edges shows many zero-conductance dips as a function of energy
(chemical potential). 
These dips are identified as resonances connected with resonant states
based on flux states which form circular-current Kekul\'{e}
patterns. It is obvious that the topology of the edges is crucial for
this phenomenon and the chirality connected with the sublattice
structure plays an important role. 
The time reversal symmetry necessary for the existence of these
resonances is violated by external magnetic fields, leading to negative
magneto resistance. 
While the structures used in the calculation might
be difficult to produce at present \cite{oshima}, our results also
suggest that transport properties of defective carbon nanotubes,
carpet-roll or  papier-m\^{a}ch\'{e}
structures \cite{zhou} could be rather different from the
transport properties of usual closed multi-wall nanotubes or SWCN which have
only weak features in the low-energy regime \cite{tamura2}. 
The present study not only 
clarifies the importance of the edges and their shapes on
transport properties,
but also indicates the importance of theoretical studies 
to explicate the interplay between 
the transport properties and
the network topology of carbon atoms.
The present numerical work provides the foresight concerning
the analysis based on a low-energy effective theory, which is beyond
the scope of this paper and will be 
presented elsewhere. 
Such an effective theory will serve as a basis for
designing carbon-based electronic devices
and for further theoretical work on effects
of impurities or electron correlation.

The authors are grateful to 
T. Enoki, C. Oshima, A. Furusaki, H. Yoshioka,
K. Kusakabe, K. Nakada, S. Okada, M. Igami and Y. Takagi
for many helpful discussions.
This work was supported partly by Grants-in-Aid for
Scientific Research No. 10309003 and 10640341 from the Ministry of
Education, Science and Culture, Japan.
K. W. acknowledges the financial support of the Japan Society for
the Promotion of Science for Young Scientist.
Numerical calculations were performed in part on VPP500 in 
Institute for Solid State Physics, 
University of Tokyo and on SX5 in Institute for Molecular Science.

\end{multicols}

\end{document}